\documentclass[11pt]{article}
\usepackage{jhep}
\usepackage[T1]{fontenc}
\usepackage{bm}
\usepackage{graphicx}
\usepackage{caption}
\usepackage{subcaption}
\usepackage[normalem]{ulem}
\usepackage{gensymb}
\usepackage[sort&compress]{natbib}
\usepackage[utf8]{inputenc}
\usepackage{bbold}
\usepackage{soul}
\usepackage[dvipsnames]{xcolor}
\usepackage{hyperref}
\hypersetup{colorlinks=false}

\preprint{UCI-HEP-TR-2019-27}

\title{\Huge QCD Baryogenesis}
\author[a]{Djuna Croon,}
\author[b]{Jessica N. Howard,}
\author[b]{Seyda Ipek,}
\author[b]{Timothy M.P. Tait}

\affiliation[a]{TRIUMF Theory Group, 4004 Wesbrook Mall, Vancouver, B.C. V6T2A3, Canada}
\affiliation[b]{Department of Physics and Astronomy, University of California, Irvine, CA 92697 USA}
\emailAdd{dcroon@triumf.ca}
\emailAdd{jnhoward@uci.edu}
\emailAdd{sipek@uci.edu}
\emailAdd{ttait@uci.edu}

\date{\today}
\abstract{
We explore a simple model which naturally explains the observed baryon asymmetry of the Universe. In this model the strong coupling is promoted to a dynamical quantity, which evolves through the vacuum expectation value of a singlet scalar field that mixes with the Higgs field. 
In the resulting cosmic history, QCD confinement and electroweak symmetry breaking initially occur simultaneously close to the weak scale.
The early confinement triggers a chemical potential between baryons and antibaryons through the interactions of the $\eta'$ meson, resulting in spontaneous baryogenesis. The electroweak sphalerons are sharply switched off after confinement and the baryon asymmetry is frozen in.
Subsequently, evolution of the Higgs vacuum expectation value (which is modified in the confined phase) triggers a relaxation to a Standard Model-like vacuum.
We identify viable regions of parameter space, and describe various experimental probes, including current and future collider constraints, and gravitational wave phenomenology. 
}

\newcommand{\SvevT}{v^T_S}

\newcommand{\beq}{\begin{equation}} 
\newcommand{\eeq}{\end{equation}}  
\newcommand{\bea}{\begin{eqnarray}}  
\newcommand{\eea}{\end{eqnarray}}

\newcommand{\fourth}{\frac{1}{4}}

\newcommand{\MeV}{\text{MeV}}
\newcommand{\GeV}{\text{GeV}}
\newcommand{\TeV}{\text{TeV}}

\newcommand{\munu}{{\mu \nu}}
\allowdisplaybreaks

\begin{document}
\maketitle

\section{Introduction}

The observed baryon asymmetry of the Universe (BAU) constitutes one of the most important open problems in modern particle physics and cosmology. Any model that explains this asymmetry must provide mechanisms that fulfill three basic (Sakharov) conditions~\cite{Sakharov:1967dj}; (i) the violation of baryon number, (ii) the violation of C and CP, (iii) reactions out of thermal equilibrium. The Standard Model (SM) does not contain the physics necessary to explain baryogenesis. Many baryogenesis models suggest new physics at scales $\gtrsim \TeV$, and realize the third Sakharov condition through the hypothesis of a first-order electroweak (EW) phase transition. 

In this paper, we take an alternative view, and ask whether baryogenesis could be a consequence of a shared cosmological history linking the electroweak and strong sectors of the SM. In particular, we study simultaneous QCD confinement and electroweak symmetry breaking (EWSB) at the weak scale. 

An analytic argument by Pisarski and Wilczek \cite{Pisarski:1983ms} suggests that QCD confinement proceeds through a first-order phase transition if the number of dynamical fermions exceeds $N_f\geq3$ at temperatures comparable to the confinement scale. 
This finding has been verified on the lattice for particular choices of $N_f$ \cite{Iwasaki:1995ij}. Confinement above the EW phase transition takes place with six massless quarks, and is thus expected to occur out of thermal equilibrium, through bubble nucleation. Through the Yukawa couplings, \emph{i.e.} $h\bar{q}u$, the quark condensate induces a tadpole term in the Higgs potential. Thus chiral symmetry breaking through QCD confinement and EWSB occur simultaneously.

If the QCD $\bar{\theta}$-angle is dynamically relaxed to zero by means of an axion, one generically expects CP violation before chiral symmetry breaking at the confinement scale. Prior to EWSB, the electroweak sphalerons are active and induce a baryonic chemical potential from the rolling axion field. This mechanism was employed to produce the BAU in~\cite{Kuzmin:1992up, Servant:2014bla,Ipek:2018lhm}. Our work completes the scenario originally proposed in \cite{Ipek:2018lhm}, in a minimal way, by realizing the relaxation to the SM vacuum after the baryon asymmetry has been frozen in. 

\begin{figure}
    \centering
    \includegraphics[width=\textwidth]{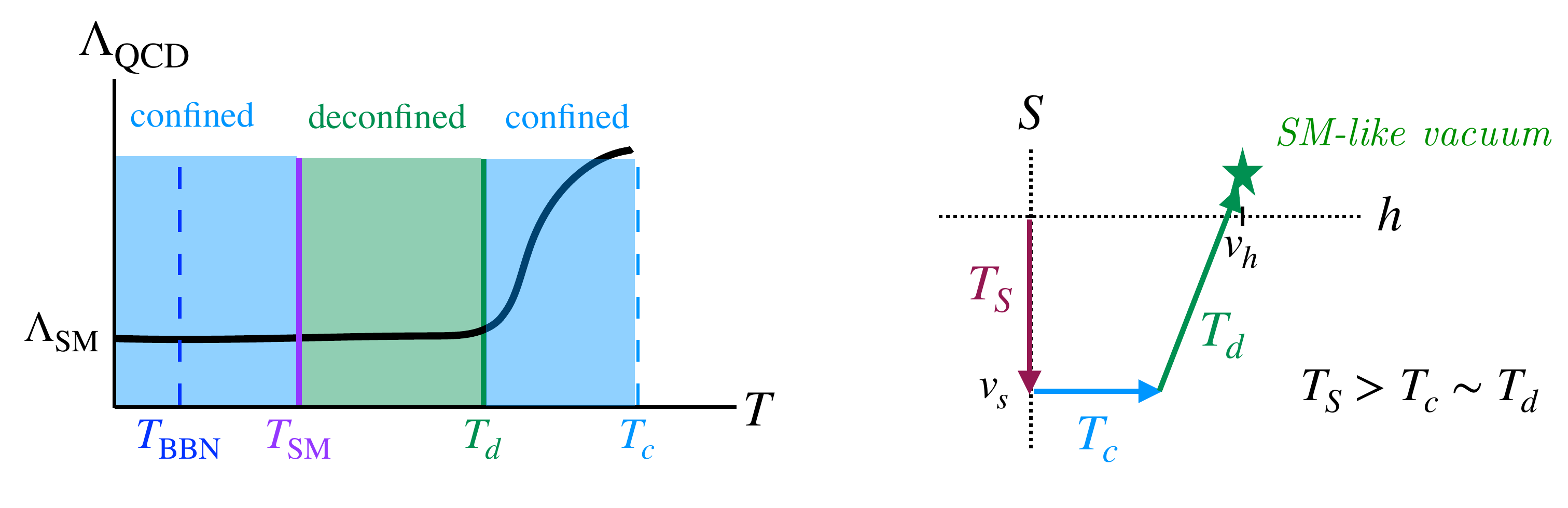}
    \caption{Schematic description of the various phases and phase transitions. At $T_c$, QCD confines at a high scale
    due to the value of $v_s$ at those temperatures.  The Higgs potential continues to evolve with temperature
    due to the plasma of electroweak
    bosons and top-flavored mesons.  At $T_d$, these corrections induce a transition to a new $v_s$ 
    (typically, but not necessarily triggering deconfinement as the QCD scale moves to its low temperature value)
    and also triggering a corresponding shift in the Higgs VEV.}
    \label{fig:symbreak}
\end{figure}

Ref.~\cite{Ipek:2018lhm} relied on a dimension-5 interaction between a real, singlet scalar field $S$ and the gluon kinetic term. When the singlet acquires a vacuum expectation value (VEV), it constitutes a contribution to the effective strong coupling and may therefore raise the QCD confinement scale. In this work we also consider the mixing between $S$ and the SM Higgs boson, and investigate the parameter space in which the EW phase transition triggers deconfinement and subsequent relaxation to the SM-like vacuum before the onset of Big Bang Nucleosynthesis (BBN), as shown schematically in Figure~\ref{fig:symbreak}.

In addition to a rather standard axion, this minimal model contains a single new degree of freedom with couplings that can be probed by future colliders. It also typically predicts a characteristic spectrum of gravitational waves, which falls within the frequency window of future space-based interferometers.

\section{A Model of Early (De)Confinement}\label{sec:model}

We consider the SM Lagrangian, with the gluon kinetic term modified to~\cite{Ipek:2018lhm},
\begin{align}
 -\fourth \left( \frac{1}{g^2_{s0}} + \frac{S}{M_\ast} \right) G^a_\munu G_a^\munu~,
 \label{eq:Lag}
\end{align}
where $G^a_\munu$ is the gluon field strength, 
$S$ is a gauge singlet real scalar field, and $g_{s0}$ represents
(after rescaling the kinetic term to canonical normalization) the
$SU(3)$ gauge coupling.
$M_\ast$ is a parameter with dimensions of energy which parameterizes a non-renormalizable interaction between $S$ and the gluons. It could be generated through the fluctuations of a radion or dilaton field, as well as by integrating out heavy vector-like $SU(3)$-charged fermions, which also couple to the scalar field $S$. In the latter case, the scale of the interaction is related to the mass of the new $SU(3)$-charged particles, $M_\ast \sim 4\pi M_Q/n_Q y_Q\alpha_s$, where $n_Q$ is the number of $SU(3)$-charged fermions with mass $M_Q$ and Yukawa coupling $y_Q$.

The scalar sector consists of the standard Higgs potential,
\begin{align}
    V(H)&= -\mu^2 |H|^2 + \lambda_h |H|^4~;
    \label{eq:LH0}
\end{align}
a potential for the $S$ field,
\begin{align}
    V(S) &= a_2 (S-S_0)^2 + a_3 (S-S_0)^3 + a_4 (S-S_0)^4,
 \end{align}   
 written in terms of the zero temperature VEV $S_0$ and three additional parameters $a_{2,3,4}$;
 and terms mixing the two scalars,
 \begin{align}
    V(H,S)&=- b_1 S |H|^2 + b_2 S^2 |H|^2,
    \label{eq:Lmix}
\end{align}
containing parameters $b_1$ and $b_2$.
The interactions in Equation~(\ref{eq:Lmix}) were presented in \cite{Ipek:2018lhm}, but neglected in the analysis for simplicity.  We show that nonzero $b_1$ and $b_2$ can play a crucial role in the dynamics, ultimately engineering the exit from the high-scale confinement phase, into the SM-like vacuum without erasing the produced baryon asymmetry at high temperatures. 

We choose parameters in the scalar potential such that the fields $H$ and $S$ have two close to degenerate local minima (including mixing terms and finite-temperature corrections) at high temperature: \textbf{i)} the high-temperature confining vacuum and \textbf{ii)} the SM-like vacuum. For an interesting region of parameter space, the high scale confining vacuum is raised as the temperature falls, resulting in the transition from it to the SM-like vacuum. 

We write the Higgs (making use of $SU(2) \times U(1)$ gauge invariance) and the singlet scalar fields as 
\begin{align}
    H=\frac{1}{\sqrt{2}}\begin{pmatrix}
                        0\\
                        v_h + \tilde{h}
                        \end{pmatrix},\quad \quad S=\frac{1}{\sqrt{2}} \left( v_s + \tilde{s} \right),
\end{align}
where $v_h \equiv \sqrt{2} \langle H\rangle$ and  $v_s \equiv \sqrt{2} \langle S \rangle$ 
are the temperature-dependent vacuum expectation values. The temperature at which $v_h$ and $v_s$ are to be considered
will usually be clear from context, and will be explicitly spelled out where necessary.  We use the notation
$v^0_h$ and $v^0_s$ to denote the zero temperature (SM-like) quantities.

\section{Thermal History} \label{sec:finiteT}

In this section, we discuss the evolution of the strong and electroweak sectors as the Universe expands and cools, based on the finite-temperature behaviour of the scalar sector, in both the confined and deconfined phases. As discussed in more detail below, the finite temperature corrections to the Higgs potential are qualitatively different in periods in which quarks and gluons are free compared to periods in which they are confined into mesons and baryons. The confinement and deconfinement phase transitions, which are both expected to be first order, are therefore described by different physics. We begin with a description of the initial high-temperature deconfined phase, $T\gg $~TeV, followed by a discussion of the physics in the confined phase.

\subsection{High Temperature and Confinement}

At temperatures above the QCD confinement scale, quarks and gluons are deconfined. 
The Higgs potential receives thermal corrections from the electroweak bosons and quarks, with the most important contribution coming from the top quark. These thermal contributions take the form
\begin{align}
   & V_{\rm gauge}(h,T) =  \sum _{i=W,Z} \frac{T^4}{2 \pi ^2}n_i  J_B\left( \frac{m_i^2}{T^2}\right) ,\quad  V_{\rm top}(H,T) =  \frac{T^4}{2 \pi ^2}n_t  J_F\left( \frac{m_i^2 }{T^2}\right),\\
&{\rm where}\quad J_{B,F}(m^2) =   \int_0^{\infty} dx\, x^2 \log \left(1\mp e^{-\sqrt{x^2+m^2}}\right),~{\rm and}~n_W = 6,~~ n_Z = 3,~~n_t = 12. \notag
\end{align}
The Higgs-dependent masses are
\begin{align}
    m^2_t = \frac{y_t^2}{2}v_h^2,~~m_W^2 = \frac{g^2}{4}v_h^2,~~ m_Z^2 =\frac{g^2+\tilde{g}^2}{4}v_h^2, 
\end{align}
where $g$ and $\tilde{g}$ are the gauge couplings of the $SU(2)$ and $U(1)$ groups of the SM, respectively, and $y_t$ is the top Yukawa coupling.

\begin{figure}
    \centering
    \includegraphics[width=.6\textwidth]{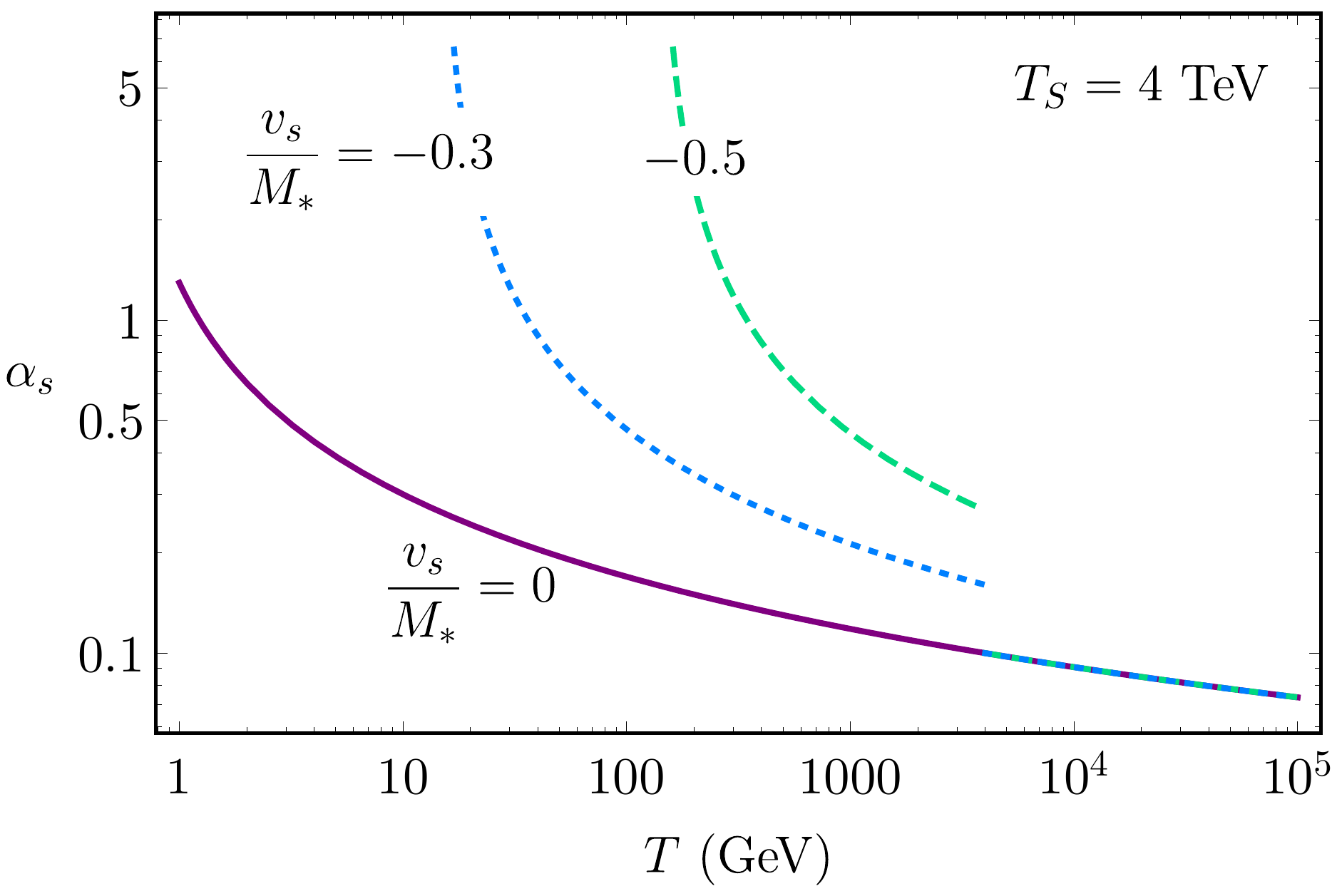}
    \caption{Evolution of the strong coupling constant with temperature in the early Universe for three different values of $v_s / M_{\ast}$, for the illustrative value $T_S=4~$TeV. Confinement takes place at temperatures for which $\alpha_s \gtrsim 1$. }
    \label{fig:alphasT}
\end{figure}

At very high temperatures, $S$ self-interactions as well as coupling to the $SU(3)$-charged particles responsible for its interaction with gluons are likely to drive its VEV $v_s$ to zero.  We choose parameters in the $S$ potential such that at some temperature $T_S \gg $~TeV, $S$ acquires a VEV, with the precise details of the value of $T_S$ and the order of this phase transition not important for our purposes. In Section~\ref{sec:benchmark} we show that $S$ self-interactions are very small and so  their thermal corrections below $T_S$ can be safely neglected. The $S$ VEV generates non-decoupling corrections to the effective strong coupling constant through the dimension-5 interaction in Equation~(\ref{eq:Lag}), which for negative $v_s$ strengthens the effective coupling strength.  At one loop, and at scale $\mu$, the effective strong coupling is 
\begin{align}
    \frac{1}{\alpha_s(\mu,\SvevT)}= \frac{33-2N_f}{12\pi}\,{\rm ln}\left(\frac{\mu^2}{\Lambda_0^2}\right)+4\pi\frac{v_s}{M_\ast},
\end{align}
where $N_f$ is the number of active quark flavors at the scale $\mu \sim T$.  Figure~\ref{fig:alphasT} shows the effective coupling as a function of temperature for the illustrative choice $T_S = 4$~TeV.

QCD confinement occurs at a temperature $T_c \simeq \Lambda_{\rm QCD}$, where
\begin{align}
    \Lambda_{\rm QCD}(v_s) = \Lambda_0 \, e^{\frac{24 \pi^2}{2 N_f - 33} \frac{v_s}{M_\ast}}. \label{eq:LQCD}
\end{align}
Here, $\Lambda_0$ is the value of the confinement scale for $v_s=0$. Correctly reproducing the strong interactions as observed at zero temperature requires, 
\begin{equation}
    \Lambda_{\rm QCD}(v_s^{0}) = \Lambda^{\rm SM}_{\rm QCD} \simeq 400~{\rm MeV},
\end{equation}
where $v_s^0$ is the zero temperature VEV for $S$.

\subsection{Evolution in the Confined Phase and Deconfinement}\label{sec:Higgsconfined}

QCD confinement results in interesting changes to the scalar potential at zero and at finite temperature.
Importantly, a vacuum expectation value for the quark condensate generates a tadpole term in the Higgs potential due to the Yukawa interactions. If confinement happens before the EW phase transition, this tadpole term triggers EW symmetry breaking. 

In the confined phase, the plasma contains mesons instead of quarks. In the proximity of the high scale QCD phase transition, we model the QCD dynamics by a nonlinear sigma model with an approximate $SU(6)_L\times SU(6)_R$ global symmetry. The Pions are embedded within a $6\times 6$ complex matrix $U (x) \equiv e^{2 i \,T^a \Pi^a (x) /f_\pi}$ which transforms as
\begin{align}
    U(x)\to L\,U(x)\,R^\dagger,
\end{align}
where $L,R$ are $SU(6)_{L,R}$ transformations respectively.

The chiral Lagrangian for mesons is  
\begin{equation}
    \mathcal{L}_{\rm chiral} = \frac{f_\pi^2}{4} \text{Tr}\left[ \partial_\mu U \partial^\mu U \right] + \kappa\, \text{Tr}\left[U \mathbf{M}\right] + \text{ H.c.}  \label{eq:Lch}
\end{equation}
where $T^a$, $a=1,\dots,35$, are the generators of $SU(6)$~\cite{doi:10.1063/1.526858}, the 
diagonal subgroup of $SU(6)_L\times SU(6)_R$ left unbroken after chiral symmetry-breaking. Since the top Yukawa is much larger than that of the other quarks, it is expected to dominate the contributions to the finite-T potential. However anticipating these mesons could be heavy enough to be Boltzmann suppressed, we keep the bottom Yukawa as well and approximate the $\mathbf{M}$ as
\begin{equation}
    \mathbf{M} \simeq \text{diag}\left(0,0,0,0,\frac{y_b v_h}{\sqrt{2}}, \frac{y_t v_h}{\sqrt{2}} \right). 
\end{equation}

The pion mass terms (and the tadpole in the Higgs potential) in the confined phase are given by
\begin{align}
   \mathcal{L}_{\rm chiral}\supset \sqrt{2}\kappa(y_t+y_b)\,v_h- \frac{2\kappa}{f_\pi^2}\Pi_a\Pi_b{\rm Tr}[\{T^a, T^b\}\mathbf{M}], \label{eq:Lchmass}
\end{align}
where $\kappa$ parameterizes the strong dynamics. The first 15 of these masses are zero in the limit where all but the top and bottom Yukawa couplings are neglected. The non-zero masses, from heaviest to lightest, are
\begin{align}
    m_{35}^2\simeq \frac{5\sqrt{2}\kappa}{3f_\pi^2}y_t v_h,~~m_{25,\dots,34}^2\simeq \frac{\sqrt{2}\kappa}{f_\pi^2}y_t v_h, ~~ m_{24}^2\simeq \frac{5\sqrt{2}\kappa}{3f_\pi^2}y_b v_h,~~m_{15,\dots,23}^2\simeq \frac{\sqrt{2}\kappa}{f_\pi^2}y_b v_h
\end{align}

The coefficient $\kappa$ is determined by matching to the SM pion mass,
\begin{align}
    m_{\pi 0}^2 = \frac{2\,\kappa_0\,(m_u+m_d)}{f_{\pi 0}^2} ~~~~~~\Longrightarrow~~~~~~ \kappa_0 = \frac{m_{\pi 0}^2 f_{\pi 0}^2}{\sqrt{2}v_h^0(y_u+y_d)}\simeq (224~{\rm MeV})^3,
\end{align}
where $m_{\pi 0}=135$~MeV is the pion mass, $f_{\pi 0}=94$~MeV is the pion decay constant and $v_h^0=246$~GeV is the 
zero temperature Higgs VEV.

During high scale confinement the effective QCD scale is modified from its SM value, 
$\Lambda_{\rm QCD}^{\rm SM}\to \Lambda_{\rm QCD}$, 
the $\kappa$ coefficient is related to its low scale analogue by\footnote{This scaling neglects a ${\cal O}(1)$ 
change due to the different number of active flavors in the two cases.}
$\kappa = \kappa_0(\Lambda_{\rm QCD}/\Lambda_{\rm QCD}^{\rm SM})^3$,
$f_{\pi 0} \to f_{\pi 0}(\Lambda_{\rm QCD}/\Lambda_{\rm QCD}^{\rm SM})$,
and thus the pion mass$^2$ scales as 
$m_{\pi0}^2\to m_{\pi 0}^2 (\Lambda_{\rm QCD}/\Lambda_{\rm QCD}^{\rm SM})(v_h/v_h^0)$, 
where $v_h$ is the Higgs VEV during high scale confinement.   Putting this together, the 
meson masses during high scale confinement are
\begin{align}
    &m_{35}^2\simeq (27~{\rm GeV})^2\left(\frac{v_h}{v_h^0}\right)\left( \frac{\Lambda_{\rm QCD}}{\Lambda_{\rm QCD}^{\rm SM}} \right),~~m_{25,\dots,34}^2\simeq (21~{\rm GeV})^2\left(\frac{v_h}{v_h^0}\right)\left( \frac{\Lambda_{\rm QCD}}{\Lambda_{\rm QCD}^{\rm SM}} \right),\notag \\
    & m_{24}^2\simeq (4~{\rm GeV})^2\left(\frac{v_h}{v_h^0}\right)\left( \frac{\Lambda_{\rm QCD}}{\Lambda_{\rm QCD}^{\rm SM}}\right),~~~~~m_{15,\dots,23}^2\simeq (3~{\rm GeV})^2\left(\frac{v_h}{v_h^0}\right)\left( \frac{\Lambda_{\rm QCD}}{\Lambda_{\rm QCD}^{\rm SM}}\right).\label{eq:mesonmass}
\end{align}


There are several novel contributions to the scalar potential in the high scale confined phase:
\begin{itemize}
\item The meson mass term in Equation~(\ref{eq:Lch}) generates a (temperature-independent) tadpole term for the Higgs:
\begin{equation}\label{eq:higgstad}
    V_{\rm tad}(v_h) = \kappa \frac{y_t}{\sqrt{2}} v_h
    \simeq - 0.0158 \, \GeV^3 ~\left(\frac{\Lambda_{\rm QCD}}{\Lambda_{\rm QCD}^{\rm SM}}\right)^3 v_h.
\end{equation}
\item The gluon condensate, $\langle GG \rangle\sim \Lambda_{\rm QCD}^4$
contributes to the $S$ potential:
\begin{align}
    V_{\rm GC}(v_s) \simeq \frac{v_s}{4M_\ast} \Lambda_{\rm QCD}^4(v_s),
\end{align}
where $\Lambda_{\rm QCD}(v_s)$ depends on $v_s$ exponentially, as described in Equation~(\ref{eq:LQCD}).
This term is typically much smaller than the other contributions to the scalar potential, however it is 
included for completeness.
\item As there are no quarks in the confined phase, the dominant thermal corrections to the 
Higgs potential are generated by top-flavored mesons (rather than top quarks):
\begin{align}
        & V_{\rm meson}(v_h,T) =  \sum _{i=15,\dots,35} \frac{T^4}{2 \pi ^2}  J_B\left( \frac{m_i^2}{T^2}  \right),
\end{align}
where $m_i, i=15,\dots 35,$ are given in Equation~(\ref{eq:mesonmass}).
\end{itemize}

We separate the complete thermal scalar potential into confined and deconfined phases, writing
\begin{align}
    V_T(v_h,v_s)=
    \begin{cases}
  V_0(v_h,v_s)+V_{\rm tad}(v_h)+V_{\rm GC}(v_s)+V_{\rm meson}(v_h,T)+V_{\rm gauge}(v_h,T)\quad ~~~{\rm (confined)}, \\
  V_0(v_h,v_s)+V_{\rm gauge}(v_h,T)+V_{\rm top}(v_h,T) \quad\quad\quad\quad\quad\quad\quad\quad\quad\quad {\rm ~~~~~(deconfined)},
    \end{cases} \label{eq:VTfull}
\end{align}
where the zero-temperature potential $V_0(v_h,v_s)=V(v_h)+V(v_s)+V(v_h,v_s)$ is given in Equations~(\ref{eq:LH0})-(\ref{eq:Lmix}).

As previously observed, QCD confinement triggers EW symmetry breaking via chiral symmetry breaking, as can be understood
from the tadpole term in the Higgs direction inducing a Higgs VEV during confinement. 
Below the confinement temperature, the Higgs potential receives thermal corrections as explained above. In this work, 
we investigate the parameter space of the $(a_i,b_i)$ coefficients in the scalar potential for which these thermal corrections 
trigger $S$ to roll (or tunnel) into a vacuum with a small positive, or zero, VEV at a deconfinement temperature $T_d$. 
In this vacuum, QCD is SM-like, \emph{i.e.} $\Lambda_{\rm QCD}=\Lambda_{\rm QCD}^{\rm SM}$. 
It is essential that this transition to SM-like QCD happens before BBN, 
e.g. $T_d>T_{\rm BBN}\sim 2$~MeV. 
If this transition happens below the EW scale, $T_d \lesssim 100~$GeV, the Higgs VEV transitions to its SM value, 
$v_h=246$~GeV at roughly the same temperature, $T_d$.   The various phases
are shown schematically in Figure~\ref{fig:symbreak}.

In the parameter region discussed below, the transition to the SM-like vacuum happens above 
$T_{\rm SM} \sim\Lambda_{\rm QCD}^{\rm SM}\sim$~ GeV, implying that
at $T_d$, quarks and gluons deconfine (again).  It is worth noting that variations could realize 
scenarios where the transition to the SM-like vacuum happens below $\Lambda_{\rm QCD}^{\rm SM}$, 
but still above $T_{\rm BBN}$, in which case QCD remains confined at all temperatures below $T_c$.

\section{Baryogenesis during QCD confinement}\label{sec:baryogenesis}

If QCD confines at a temperature when the Higgs VEV is zero, \emph{i.e.} quarks are massless, the phase transition is 
expected to be first order \cite{PhysRevD.29.338} and proceeds through bubble nucleation. This first-order phase 
transition, combined with an axion solution to the strong CP problem, results in a novel baryogenesis mechanism. 
In \cite{Ipek:2018lhm} this phase transition was imagined to occur at $T>T_{\rm EW}$ such that Higgs VEV is expected to be zero because of the SM thermal corrections. 

\begin{figure}
    \centering
    \includegraphics[width=0.98\textwidth]{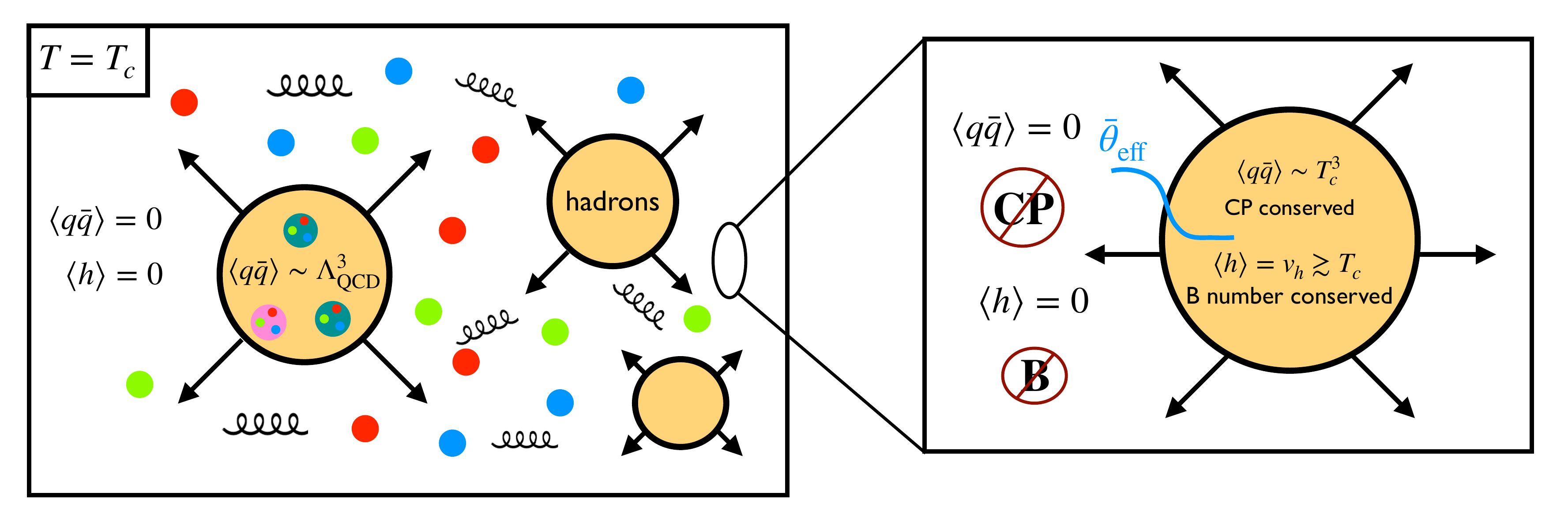}
    \caption{Schematic description of QCD confinement, bubble nucleation and the baryon asymmetry generation. 
    QCD confines at a temperature $T_c$ via a first-order phase transition, for which the Higgs VEV is zero outside the bubbles, 
    which means baryon number violation is efficient via EW sphalerons. If there is a QCD axion, there is also typically 
    large CP violation due to the uncancelled $\bar{\theta}$ angle, which shuts off in the confined phase.}
    \label{fig:QCDBG}
\end{figure}

In this work, we highlight a scenario in which although the QCD confinement happens at 
$T_c \lesssim 100$~GeV, the Higgs VEV before confinement is zero due to the extended scalar sector. As long as the 
EW symmetry is unbroken, sphalerons are active and baryon number is efficiently violated. 
This is the case outside of the bubbles of confined phase. Inside, the QCD confinement triggers EW symmetry 
breaking and sphalerons are inoperative, thus preserving any baryon asymmetry.  The need for CP violation can be accounted for if there is large CP violation from the uncancelled strong phase before the axion rolls to the minimum of its 
potential~\cite{Kuzmin:1992up, Servant:2014bla,Ipek:2018lhm}. (Note that this is essentially a spontaneous baryogenesis mechanism~\cite{Cohen:1988kt}.)
Figure~\ref{fig:QCDBG} provides a schematic description. 

In this section we summarize the mechanism for the generation of the baryon asymmetry.
The axion is driven to its minimum after confinement occurs, dynamically solving the strong CP problem. 
However, as it rolls, there is an uncancelled effective $\bar{\theta}$. This nonzero $\bar{\theta}$ induces a $G\tilde{G}$ condensate,
which couples to the baryon current via the pseudoscalar $\eta'$ meson,
whose mass scales like $m_{\eta'} \sim \Lambda_{\rm QCD}$. 
At energies below this mass, its residual effects are described by the effective Lagrangian~\cite{Kuzmin:1992up}
\begin{align}
\mathcal{L}_{\rm eff} \simeq \frac{10}{f_{\pi}^2 m_{\eta'}^2}\frac{\alpha_s}{8\pi} \langle G\widetilde{G} \rangle 
\frac{\alpha_w}{8\pi}  W\widetilde{W},
\end{align}
where $W$ ($\widetilde{W}$) is the $SU(2)_W$ (dual) field strength and
\begin{align}
\frac{\alpha_s}{8\pi} \langle G\widetilde{G} \rangle = m_a^2(T)f_a^2\sin\bar{\theta}.
\end{align}
Here $m_a(T)$ is the temperature-dependent axion mass (we ignore the temperature dependence of $\bar{\theta}$ for simplicity). This temperature dependence has been calculated analytically~\cite{PhysRevD.33.889} and by lattice studies~\cite{WANTZ2010110} at various temperature regimes. For our model the relevant temperature dependence of the axion mass can be summarized as 
\begin{align}
    m_a^2(T)f_a^2 \simeq \begin{cases}
                    m_\pi^2 f_\pi^2\,\bar{m}\quad\quad\quad\quad\quad\quad\quad\quad\quad~~ T<\Lambda_{\rm QCD}, \\
                    \zeta\, m_\pi^2 f_\pi^2\,\bar{m}  \left(\frac{\Lambda_{\rm QCD}}{T}\right)^n~~~T>\Lambda_{\rm QCD}\sim T_{\rm EWSB},
                    \end{cases} \label{eq:maT}
\end{align}
where $m_\pi^2 f_\pi^2\simeq m_{\pi 0}^2 f_{\pi 0}^2(v_h/v_h^0)(\Lambda_{\rm QCD}/\Lambda_{\rm QCD}^{\rm SM})^3$ and $\bar{m}=\sqrt{m_u m_d}/(m_u+m_d)\simeq 0.5$ (see, \emph{e.g.}, Equations (18)-(23) in \cite{PhysRevD.45.466}). The parameter $\zeta$ and the exponent $n$ represents different temperature regimes above the QCD confinement scale and depends on the number of light flavors. 

\begin{figure}
    \centering
    \includegraphics[width = .5\textwidth]{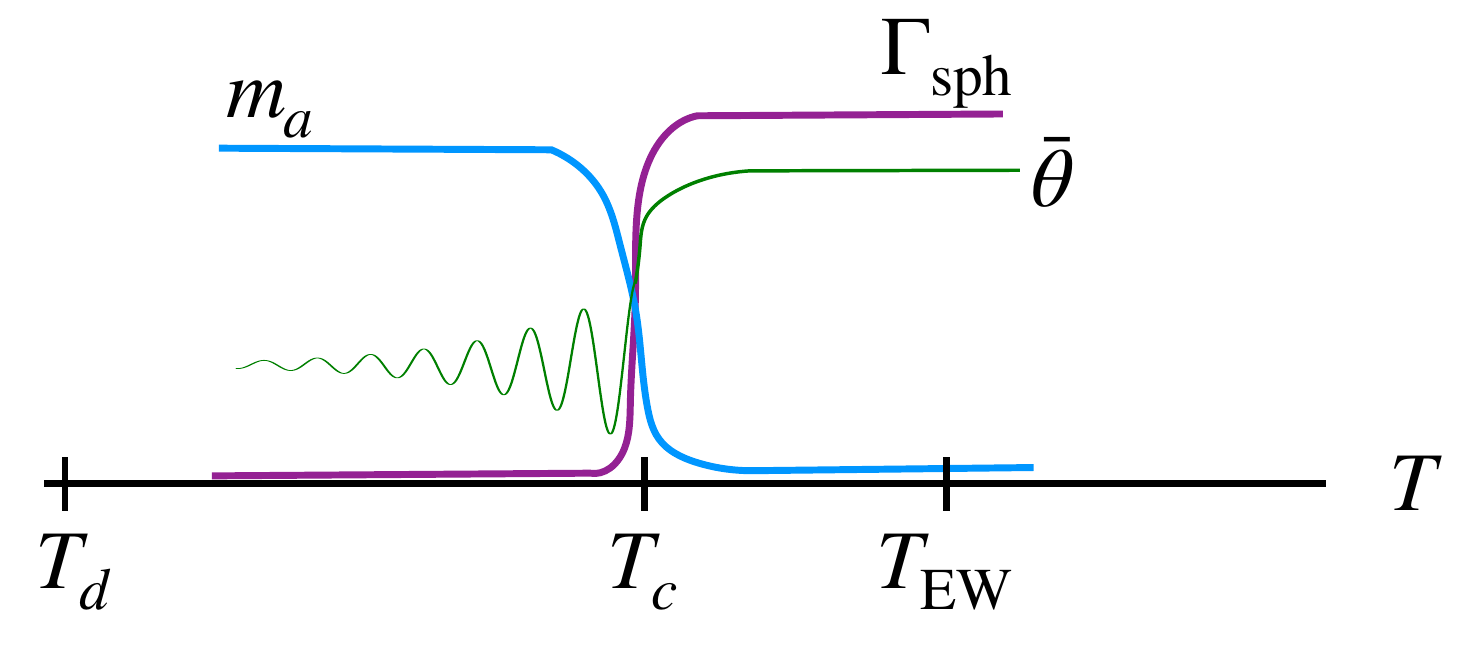}
    \caption{Schematic description of various quantities that are involved in producing the baryon asymmetry. During QCD confinement at $T_c$, changing axion mass generates a chemical potential between baryons and antibaryons. Sphalerons turn off after QCD confinement because EW symmetry is broken.}
    \label{fig:baryorates}
\end{figure}

Since $W \widetilde{W}$ is connected to the baryon current density $j_B^\mu$ through the anomaly equation, 
$\partial_\mu j_B^\mu = (\alpha_W / 8 \pi){\rm Tr}[ W \widetilde{W}$], the $G\tilde{G}$ condensate generates an effective chemical potential~\cite{Cohen:1987vi,Cohen:1988kt,Co:2019wyp} 
for baryons given by
\begin{align}
    \mu = \frac{10}{f_\pi^2 m_{\eta'}^2}\sin\bar{\theta}\frac{d}{dt}\left[m_a^2(T) f_a^2\right].
\end{align}
Here we note that $\eta'$ acquires all of its mass from chiral symmetry breaking and as such $m_{\eta'}^2 f_\pi^2 \propto m_{\eta'0}^2 f_{\pi 0}^2 (\Lambda_{\rm QCD}/\Lambda_{\rm QCD}^{\rm SM})^4$.

The change in baryon number is given by 
\begin{align}
    n_B = \int_{T_i}^{T_f} dt \frac{\Gamma_{\rm sph}(T)}{T}\mu,
\end{align}
where $\Gamma_{\rm sph}\sim 25\alpha_w^5 T_{\rm sph}^4$~\cite{Moore:2010jd,DOnofrio:2014rug} is the EW sphaleron rate in thermal equilibrium. 

The baryon-to-entropy ratio can be approximated as 
\begin{align}
\eta =\frac{n_B}{s}\simeq \frac{45\times 125}{2\pi^2 g_*(T_{\rm reh})}\alpha_w^5\sin\bar{\theta}\frac{\Delta\left[m_a^2(T)f_a^2\right]_{T_c}}{m_{\eta'}^2 f_\pi^2}  \left(\frac{T_{\rm sph}}{T_{\rm reh}}\right)^3,
\end{align}
where $T_{\rm reh}$ is the reheat temperature 
at the end of the QCD/EW phase transition, and $g_*\simeq 53$ counts the number of relativistic degrees of freedom in equilibrium~\footnote{Light degrees of freedom are: 26 mesons, gluons, photon, the singlet and leptons.} at that 
time. From Equation~(\ref{eq:maT}) the change in the axion mass over the confinement temperature is $\Delta\left[m_a^2(T)f_a^2\right]_{T_c} \simeq m_\pi^2 f_\pi^2\,\bar{m}$. This gives
\begin{align}
    \eta &\simeq 4.4\times 10^{-9}\sin\bar{\theta}\left(\frac{v_h}{v_h^0}\right)\left(\frac{\Lambda_{\rm QCD}^{\rm SM}}{\Lambda_{\rm QCD}}\right) \left(\frac{T_{\rm sph}}{T_{\rm reh}}\right)^3 \simeq 10^{-11}\sin\bar{\theta}\left(\frac{v_h}{\Lambda_{\rm QCD}}\right)\left(\frac{T_{\rm sph}}{T_{\rm reh}}\right)^3,
\end{align}
which is to be compared with the Planck measurement \cite{Ade:2015xua}, 
\begin{align}
\eta_{\rm obs}=(8.59\pm 0.11)\times 10^{-11}.
\end{align}

In the benchmark scenarios we study below, $v_h/\Lambda_{\rm QCD}\sim 1-4$ (we discuss below why this ratio should be larger than 1). 
The numerical proximity of the above estimate of the baryon asymmetry to the observed value suggests $\sin\bar{\theta}\simeq1$ before confinement, and points at a QCD confinement scale below the EW scale. 

Interestingly, in models of \emph{cold baryogenesis}, \emph{e.g.}~\cite{GarciaBellido:1999sv}, lattice simulations shows that $T_{\rm sph}/T_{\rm reh}\sim 40$~\cite{Tranberg:2006dg}. In these models the Higgs-mass term is modified to be $\mu_{\rm eff}^2=\mu^2 - b\, S^2$ (in our case this becomes $\mu_{\rm eff}^2 = \mu^2+b_1 S -b_2 S^2$, with a negative $S$ VEV at high temperature). While the scalar field $S$ moves along its potential, the effective Higgs mass term changes sign. When the mass term crosses zero, long-wavelength gauge configurations are produced out of thermal equilibrium and these source sphaleron transitions. Another class of models that can produce a similarly larger-than-eqilibrium sphaleron rate is where the Higgs vacuum is stuck in the metastable, zero-VEV vacuum until temperatures below the EW scale. The decay of the false vacuum can be triggered by QCD confinement, as in \cite{WITTEN1981477}, and could generate the out-of-equilibrium sphaleron configurations, again causing the sphaleron temperature to be larger than the equilibrium temperature. Our model is a combination of these two scenarios and we expect that $T_{\rm sph}/T_{\rm reh}>1$. However, a more detailed study is needed, with lattice input, and will be carried out in future work.

Successfully realizing this picture for baryogenesis requires: 
\begin{itemize}
\item The sphalerons must be active at the time of the high scale QCD confinement, requiring 
 $v_h^{T>T_c} = 0$.  In Ref.~\cite{Ipek:2018lhm}, this was trivially satisfied by choosing $T_c>T_{\rm EW}$. 
 However, we find additional parameter regions where $T_c <  T_{\rm EW}$, but for which the extended scalar sector
 delays the onset of EW symmetry breaking until confinement at $T_c$.
\item There is the danger of washing out the generated baryon asymmetry if the sphalerons remain sufficiently active
 inside the bubbles of the confined phase.  
 Provided the Higgs VEV inside the bubbles of the confined phase
 satisfies $v_h^{T_c} / T_c\gtrsim 1$, this is not a concern. 
\end{itemize} 

\section{Benchmark Parameter Space} \label{sec:benchmark}

In this section we explore benchmark regions of parameter space numerically. In total, there are ten parameters:
\begin{itemize}
    \item two parameters in the Higgs potential: $ \mu$ and $\lambda$;
    \item four parameters in the singlet potential $V(S)$: $S_0$ and $a_{2,3,4}$;
    \item two scalar mixing parameters: $b_1$ and $b_2$;
    \item the scale of the $S$ interaction with gluons, $M_\ast$; and
    \item $\Lambda_0$, the confinement scale for $v_s=0$.
\end{itemize}
We fix two of these parameters ($\mu$ and $\lambda$) by imposing that the SM vacuum is realized for the SM Higgs VEV,
$v_h^0 = 246 \, \GeV$ and contains a Higgs-like mass eigenstate  of mass $m_h = 125 \, \GeV$. 
Reproducing the correct $\Lambda_{\rm QCD}^{\rm SM} \simeq 400\, \MeV$ fixes a combination of
$S_0$ and $\Lambda_0$. In the following we choose $\Lambda_0 = 500$~MeV which fixes $S_0$ appropriately based on the remaining parameters.

\begin{table}[t]
\centering
\begin{tabular}{lllll} &$M_*$& $a_2/\GeV^2$ & $a_3/\GeV$ & $a_4$ \\ \hline
 1. &$1.5 \,\TeV$ & $380 $ & $9.9 \times 10^{-1}$ & $6.3\times 10^{-4}$ \\ 
 2. &$3 \,\TeV$  & $108 $ & $1.5 \times 10^{-1}$ & $5.1\times 10^{-5}$ \\
 3. &$3 \,\TeV$  & $44.2$ & $6.14 \times 10^{-2}$ & $2.1 \times 10^{-5}$ \\ 
 4. &$5 \,\TeV$ & $38.9 $ & $3.24 \times 10^{-2}$ & $6.6\times 10^{-6}$  \\ 
 5. &$10 \,\TeV$  & $9.72 $ & $4.05 \times 10^{-3}$ & $4.1\times 10^{-7}$  \\ 
 6. &$10 \,\TeV$  & $4.92 $ & $2.27 \times 10^{-3}$ & $2.6\times 10^{-7}$ 
\end{tabular}
\caption{Six benchmark parameter choices, described in the text.}\label{table:bench}
\end{table} 

\begin{figure}
    \centering
    \hspace*{-0.5cm}
    \includegraphics[width=1.05\textwidth]{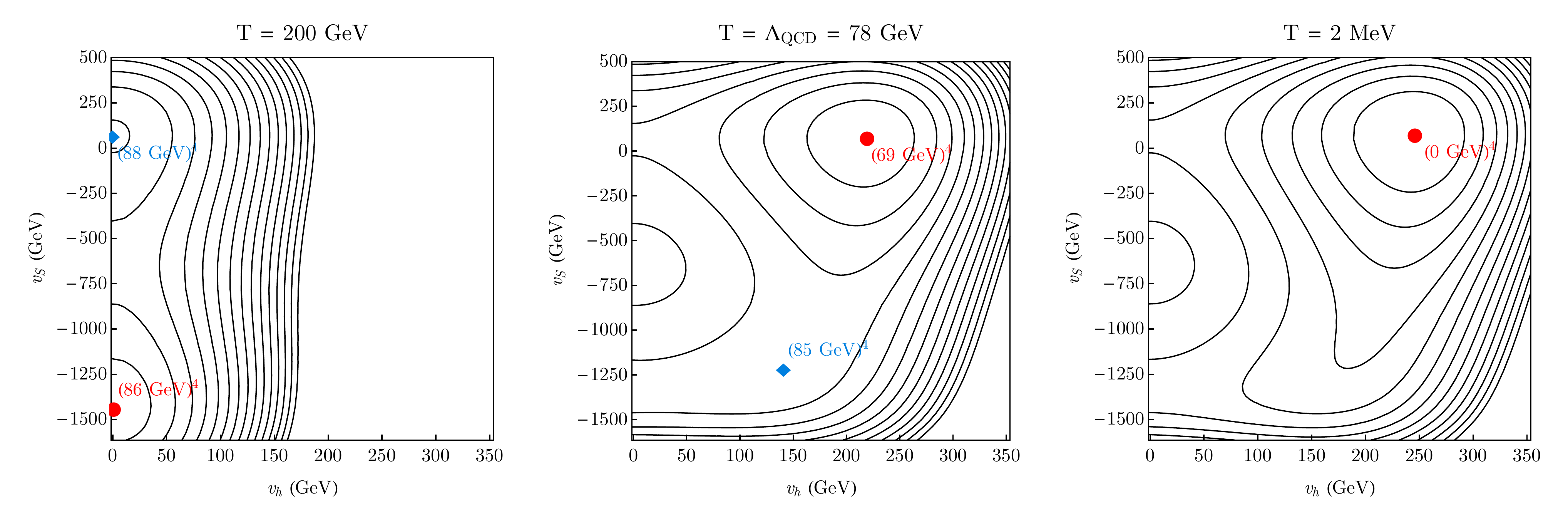}
    \caption{Contours of constant potential for the benchmark point 2, with $b_1 = 0.2 \,\GeV$, 
    and $b_2= 1. \times 10^{-3}$ at three indicated temperatures.  In each panel, 
    the red dot marks the global minimum at that temperature, and the blue diamond a local minimum. We point out that at $T=T_c$ the global minimum is separated from the local minimum by a barrier and tunneling is not efficient enough. At a lower temperature, $T_d$, this barrier disappears and the system rolls to the SM-like vacuum.}
    \label{fig:bm1potential}
\end{figure}

\begin{figure}
    \centering
    \includegraphics[width=0.48\textwidth]{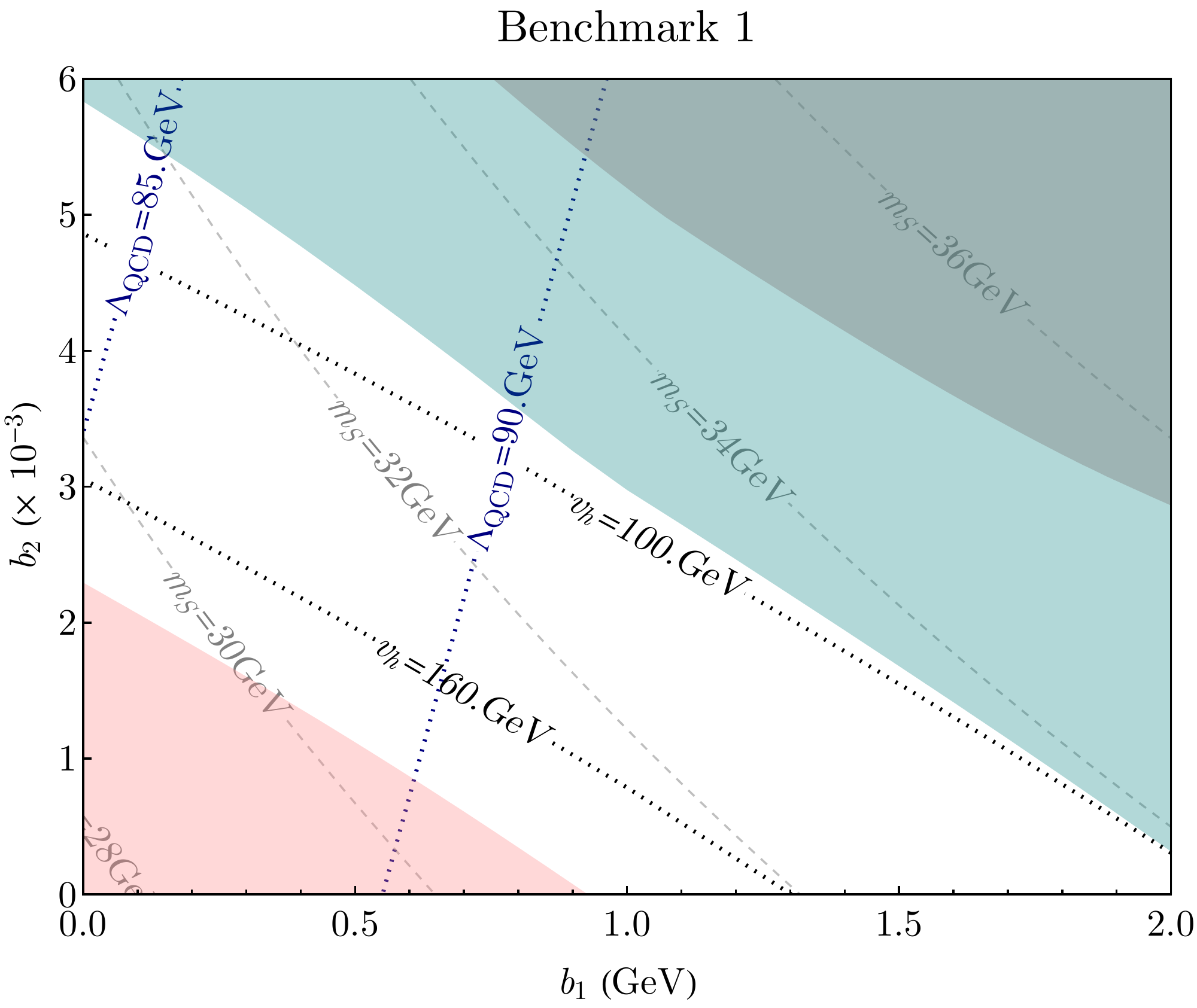}
    \includegraphics[width=0.48\textwidth]{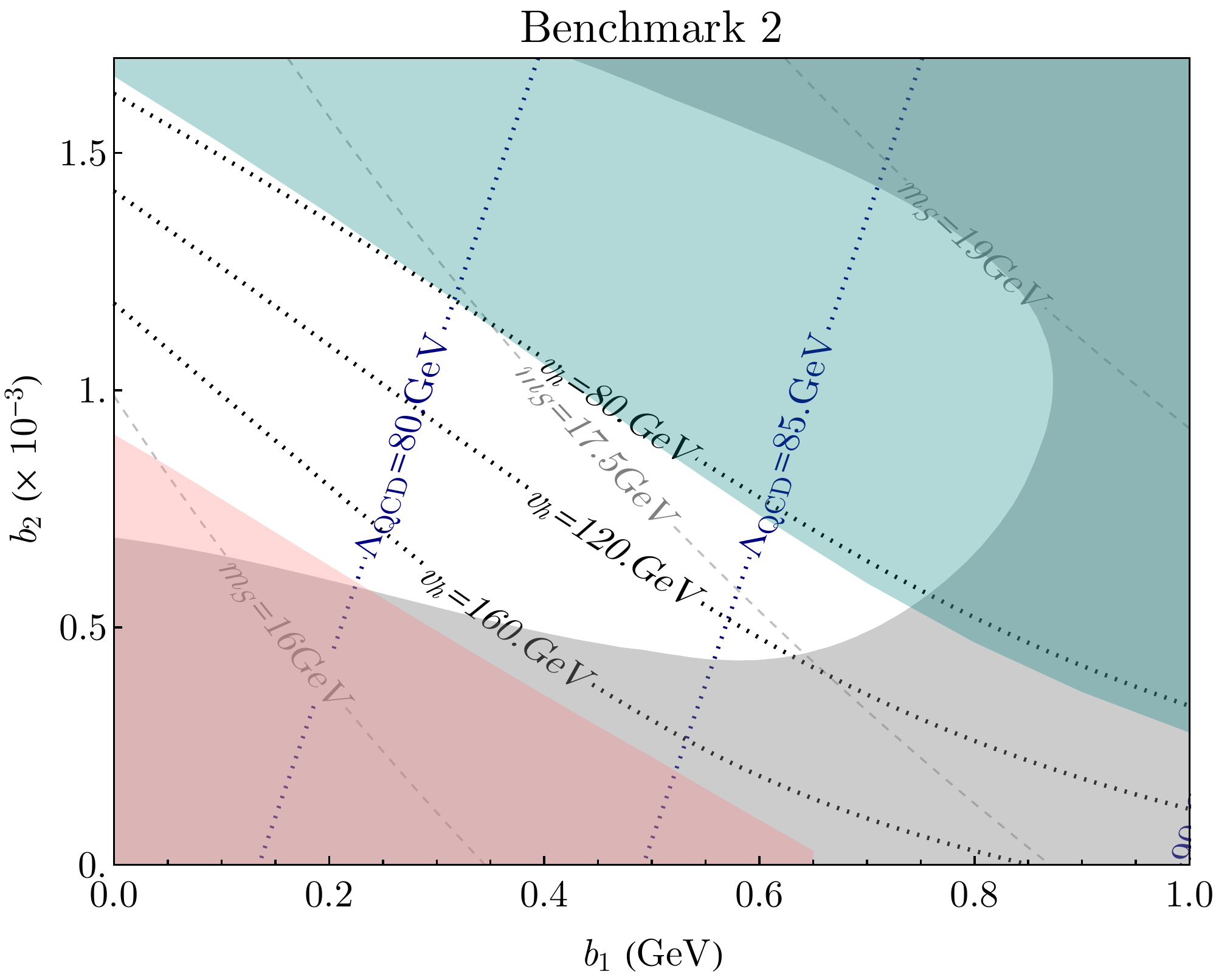}
    \caption{Parameter space in the $b_1$-$b_2$ plane for benchmark 1 (left panel) and 2 (right panel). 
    In the blue-shaded, upper-right region, $v_h<T$ after confinement, risking washout of the baryon asymmetry. 
    In the pink-shaded, lower-left region, $v_h>T$ before confinement, and the sphalerons are inactive at 
    $T \sim \Lambda_{\rm QCD}$. 
    Gray shading marks the region in which tunneling to the SM-vacuum is highly suppressed,     and inaccessible before $T_{\rm BBN} = 2$~MeV. }
    \label{fig:bm1}
\end{figure}

This leaves six parameters: $M_*$; $a_2$, $a_3$, and $a_4$ characterizing the remainder of $V(S)$, 
and the two Higgs-$S$ couplings $b_1$ and $b_2$.  
Given the large dimension of the parameter space, it is not practical to scan over all of them, and thus
we define six benchmark scenarios for $M_\ast$ 
(ranging from $M_\ast = 1.5$~TeV to $M_\ast = 10$~TeV), and the $a_i$
in Table~\ref{table:bench}, and scan over $b_1$ and $b_2$. Our results are shown in Figures~\ref{fig:bm1potential}-\ref{fig:bm1}, which are based on a $25\times25$ linearly spaced parameter scan, using a cubic regression method to smooth the contours.

In Figure~\ref{fig:bm1potential}, we show a contour plot of the scalar potential (in the $v_h$-$v_s$ plane) for Benchmark point 2 with $b_1 = 0.7 \,\GeV$, and $b_2 = 1.0 \times 10^{-3}$ for three different temperatures: $T = 200$~GeV~$> T_c$, $T = 85$~GeV = $\Lambda_{\rm QCD}$, and $T = 2$~MeV, illustrating the salient points leading to a cosmological evolution which successfully realizes the thermal history described above. At high temperatures, finite-temperature corrections in the Higgs direction ensure $v_h = 0$, and $v_s$ takes the (negative) value defined by $S_0$, eventually leading to high scale confinement when $T \sim \Lambda_{\rm QCD}(v_s)$. Before confinement, but after $T\sim T_{\rm EW}$, the SM-like point becomes the global minimum. However, it is sepearated from $v_h=0$ vacuum by a barrier.  Typically, the large distance in field space between the two vacua implies that tunneling  is expected to be strongly suppressed before confinement, as the Euclidean bounce action scales as  $S_E/T \propto \left(\Delta \phi_i / \Delta V(\phi_i) \right)^4 \sim 10^{-8}$ (see e.g. \cite{Croon:2018erz,Croon:2018kqn}). Hence the universe is in a metastable state. Confinement triggers a change in the degrees of freedom contributing to the effective potential in the Higgs direction, as described in Section~ \ref{sec:Higgsconfined}, which in turn quickly shifts the minimum to non-zero $v_h$ due to the tadpole term from chiral symmetry-breaking. Note that this minimum may still be a local minimum, separated from the true, SM-like, vacuum by a potential barrier. As the temperature drops further, the thermal corrections become less important while the mixing terms governed by  $b_1$ and $b_2$ take over. Subsequently the mixing terms lift the potential and eventually the field can roll to the SM-like vacuum, instead of tunneling.

Figure~\ref{fig:bm1} summarizes the allowed parameter space in the $b_1$-$b_2$ plane for benchmarks 1 and 2.
The black dotted lines indicate contours of the labelled values for $v_h$ at $T_c$, and blue dotted lines show contours
for the values of $\Lambda_{\rm QCD}$.
The pink-shaded region corresponds to the parameter space in which sphalerons are not active at the time of QCD confinement,
$v_h \neq 0$ right before $T_c$, whereas the blue-shaded region corresponds to the parameter space in which $v_h / T_c  < 1$,
risking that the generated baryon asymmetry will be washed out.  Grey shading indicates points in which transitioning to
the SM-like vacuum is via suppressed tunneling, and would not occur before BBN.  The white (unshaded) region thus
allows one to successfully realize baryogenesis as described above.  Also shown for reference are contours of fixed mass of the
singlet $S$ at zero temperature (see Section~\ref{sec:discussion}, below).

The benchmark studies reveal a few features which are likely to be fairly generic:
\begin{itemize}
    \item Larger choices of $M_*$ require larger values of $v_s$ in the confining vacuum to obtain the same $\Lambda_{\rm QCD}$ (see Equation (\ref{eq:LQCD})), 
    which in turn corresponds to a larger distance in singlet-field space between the vacuum during high scale
    confinement and  the SM-like vacuum.  The need to transition to the SM-like vacuum before BBN then implies
    smaller values of the couplings $b_1$, $b_2$, translating into smaller singlet masses. 
    \item In order for the Higgs VEV to successfully trigger deconfinement, 
    it is necessary that the potential be EW scale in the $S$ direction, at least for a distance in field space 
    $\Delta S \sim \mathcal{O}(M_*)$.  That implies that the parameters $a_i$ should be inversely 
    correlated with the scale $M_*$, as was engineered for the benchmark points.
    \item Note that all of the benchmark models have a vacuum energy which is always subdominant to the energy of the 
    SM radiation bath.
\end{itemize}

\section{Experimental Constraints and Prospects} \label{sec:discussion}

The signature property of the kind of modification of the QCD coupling described here is the existence of scalar excitations
of the $S$ field, which couple to gluons and mixes with the Higgs boson through the interactions $b_1$ and $b_2$.
In this section we discuss current constraints and prospects for future searches for the viable regions of $b_1$ and $b_2$
corresponding to the six benchmark models defined in Table~\ref{table:bench}.

We denote the mass eigenstates by $h$ and $s$.  They are related to the gauge basis by an orthogonal transformation,
\begin{align}
    \begin{pmatrix}
    h\\
    s
    \end{pmatrix} = \begin{pmatrix}
                    \cos\theta && - \sin\theta \\
                    \sin\theta &&\cos\theta
                    \end{pmatrix}  \begin{pmatrix}
    \tilde{h}\\
    \tilde{s}
    \end{pmatrix}.
\end{align}
which is itself a consequence of the scalar potential at zero temperature.  
Details are presented in Appendix~\ref{sec:smixing}.
We order the eigenvalues such that
$h$ is the mostly SM Higgs-like state, with a mass close to 125 GeV, and $\theta \ll 1$.  
In this regime, the dominant contribution to the $s$ mass is through the mixing with the Higs, and is typically of 
${\cal O}(10~{\rm GeV})$.
Experimental
measurements restrict the mass of the $s$, the mixing angle $\theta$,
and the scale\footnote{Technically, the $M_\ast$ relevant for cosmology is at scales of order $\Lambda_{\rm QCD}$, whereas
the quantity relevant for phenomenological probes depends on the scale of the observable in question.  We neglect this
subtlety.}
$M_\ast$.

\subsection{Probing the $(S/M_\ast)GG$ interaction}

The scale of the singlet--gluon interactions, $M_\ast$, can be constrained at hadron colliders in a model-independent way. 
Singlet production at the LHC is dominated by gluon fusion, and its decays are also mainly back into gluons with
tree level partial decay width
\begin{align}
    \Gamma(s \to gg) \simeq \frac{m_s^3}{8\pi M_\ast^2}~.
\end{align}
Given the tiny mixing parameters in the benchmark scenarios, this is always the dominant decay mode (though
for some parameters, decays into quarks can be comparable, see Figure~\ref{fig:partial}), and thus
this partial width characterizes the $s$ lifetime.  In the benchmark models, 
the longest lived singlet has $m_s \sim 5$~GeV and $M_\ast = 10$~TeV, which results in a prompt decay length of 
$c\tau \sim 10^{-7}$~cm. 

The dijet resonance search by ATLAS~\cite{ATLAS:2019bov} constrain $M_\ast \gtrsim 4-15~$TeV for singlet 
masses 2-4~TeV \cite{Danielsson:2019ftq}. However, the singlet masses of prime interest are 
$\mathcal{O}(10~{\rm GeV})$, and in this regime non-resonant searches are more useful. 
In~\cite{Gavela:2019cmq} such searches were reinterpreted for axion-like particles (ALPs). In contrast to a generic ALP, 
the singlet here dominantly couples to the gluons rather than to weak or hypercharge bosons. 
Hence, most of the constraints, \emph{e.g.} from decays to photons, are not applicable. 
However the CMS dijet angular distribution~\cite{Sirunyan:2018wcm} requires $M_\ast \gtrsim 3~$TeV, 
independent of the ALP mass provided it is $\lesssim100$~GeV.
The scalar $s$ has a different structure than the pseudo-scalar ALP in its coupling to gauge boson polarizations,
and so this limit is likely to be modified at ${\cal O}(1)$. 

Though model-dependent, there may be additional searches for $SU(3)$-charged
particles responsible for generating the $S$ coupling to gluons.  One possibility
is through a loop of heavy vector-like quarks, in which 
case $M_\ast \sim M_{\rm VLQ}$.  There are various model-dependent LHC searches for vector-like quarks, with particular emphasis on searches for top partners. 
Searches for pair-produced vector-like quarks mixing with the SM top quark 
exclude their masses below $\sim 1$~TeV~\cite{Aaboud:2017qpr,Aaboud:2017zfn} 
whereas single production constraints go up to $1.4$~TeV~\cite{Sirunyan:2017tfc},
depending on the mixing angle. 
Bounds are typically somewhat weaker for vector-like quarks mixing with lighter
SM quarks.

\subsection{Scalar mixing}

The requirement that the transition to the SM happen before BBN points to the mixing parameters 
$b_1\lesssim$~a few GeV and $b_2\sim\mathcal{O}(10^{-5}-10^{-3})$. 
In the benchmark scenarios, the scalar mass is $\mathcal{O}$(1-10~GeV). 
Together, these parameters allow a small mixing between the singlet and the Higgs, $\sin\theta\sim 10^{-4} -10^{-2}$.
This is below the current sensitivity of the LHC to the properties of the Higgs boson \cite{Khachatryan:2016vau}.

\begin{figure}
    \centering
    \includegraphics[width=0.48\textwidth]{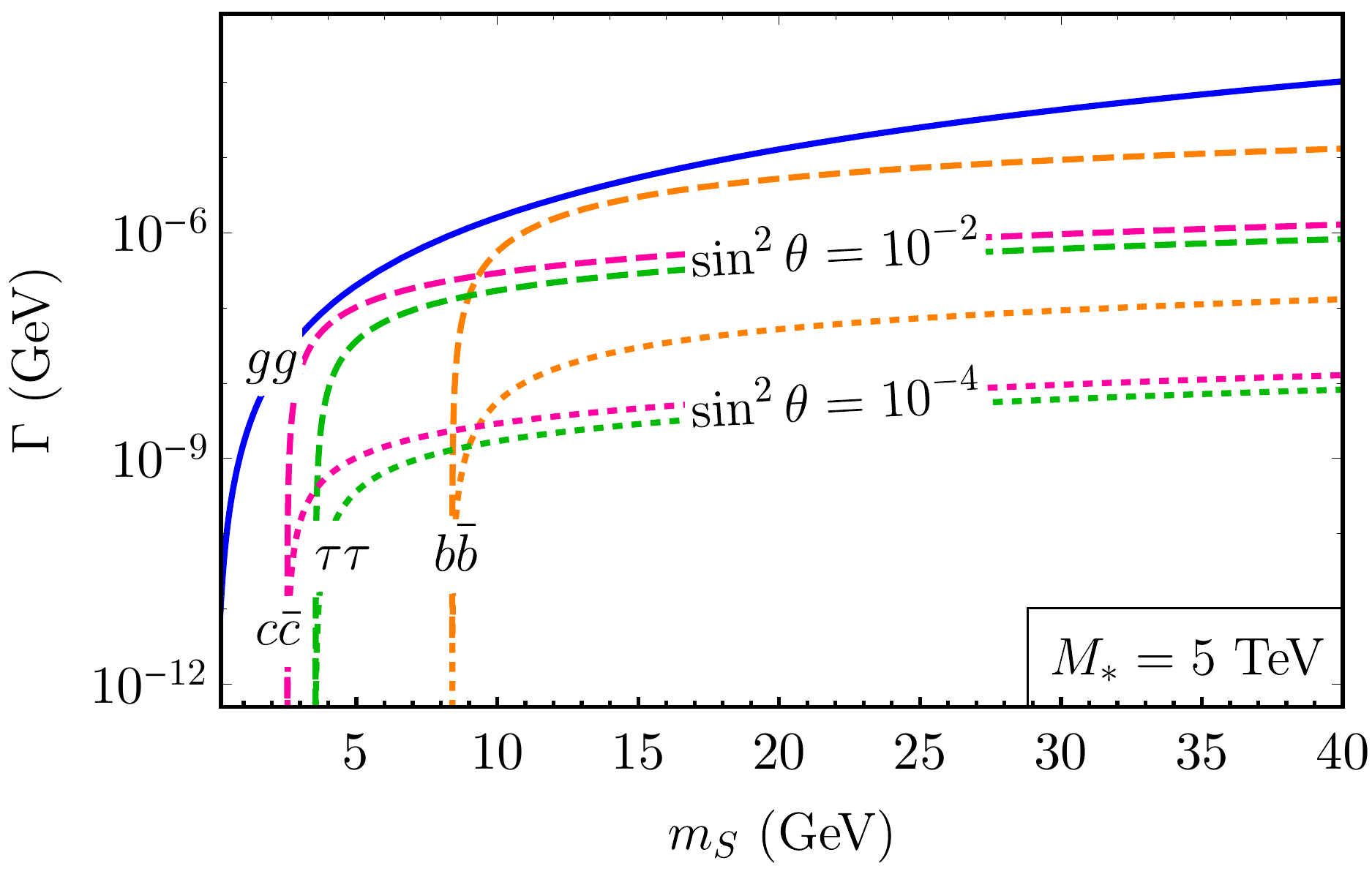}
    \includegraphics[width=0.48\textwidth]{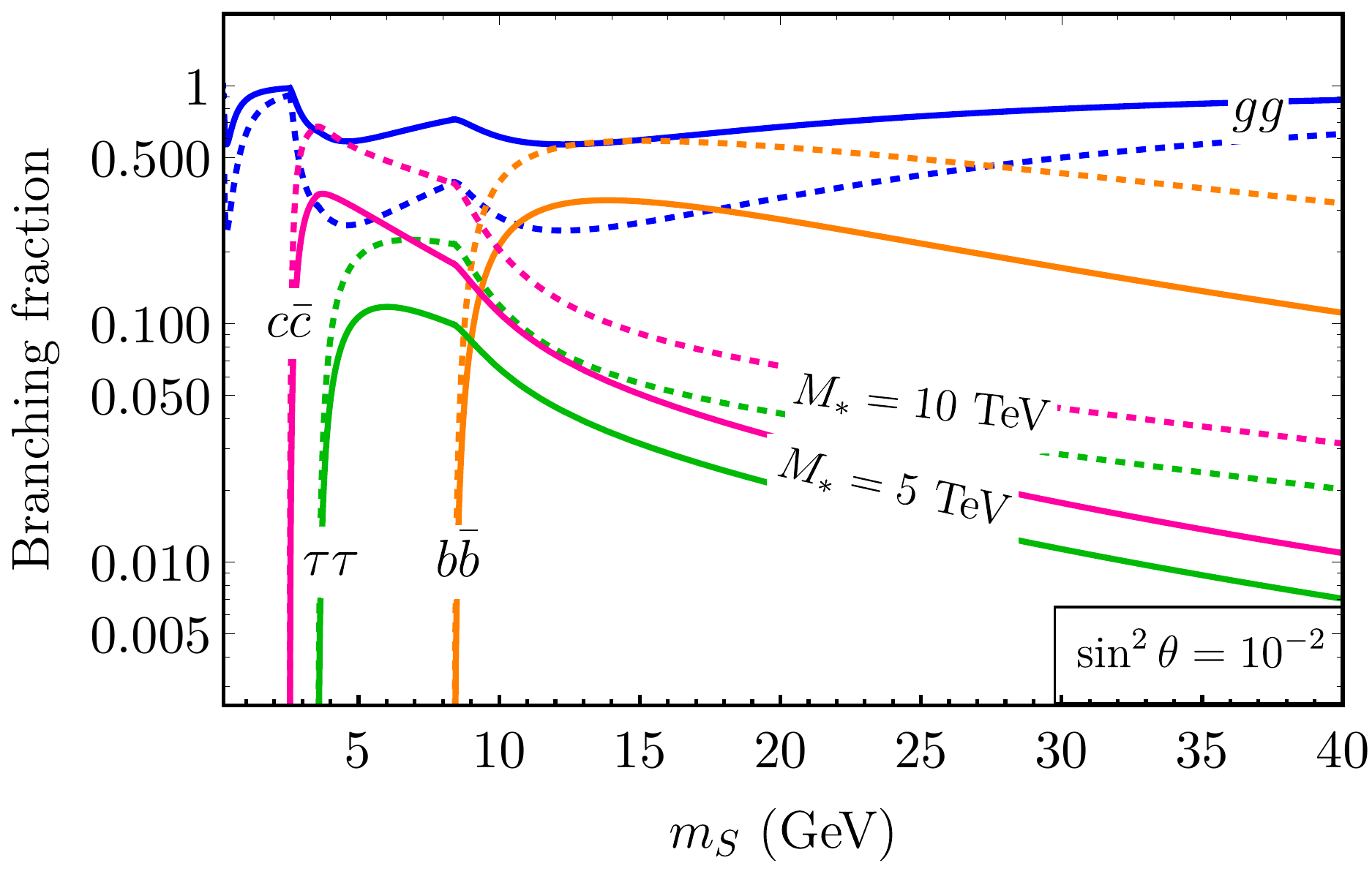}
    \caption{\textbf{(Left:)} Partial decay width of $s$ into gluons, $b\bar{b}, c\bar{c}$ and $\tau^+\tau^-$ final states. 
    For reference we show two mixing angles, $\sin^2\theta =10^{-2}$ (dashed) and $10^{4}$ (dotted). 
    We set the scale of the scalar-gluon interaction at $M_\ast=5$~TeV. \textbf{(Right:)} Branching fractions into final states of 
    gluons,$b\bar{b}, c\bar{c}$ and $\tau^+\tau^-$ for $M_\ast=5$~TeV (solid) and 
    10 TeV (dotted) for $\sin^2\theta = 10^{-2}$. }
    \label{fig:partial}
\end{figure}

The singlet decays into SM fermions $f$ via its mixing with the Higgs boson. The partial decay width into $f \bar{f}$ is given 
at tree level by
\begin{align}
    \Gamma(s \to f\bar{f}) \simeq \frac{N_c\,y_f^2\sin^2\theta\, m_S}{8\pi}\left(1-\frac{4m_f^2}{m_S^2}\right)^{3/2},
\end{align}
where $N_c=3$ for quarks and 1 for leptons.  In Figure~\ref{fig:partial} we show the partial decay widths into 
gluon and fermion final states, and the branching ratios as a function of $m_S$ 
for representative values of $\sin \theta$ and $M_\ast$ from the benchmark models. 

\begin{figure}
    \centering
    \includegraphics[width=0.7\textwidth]{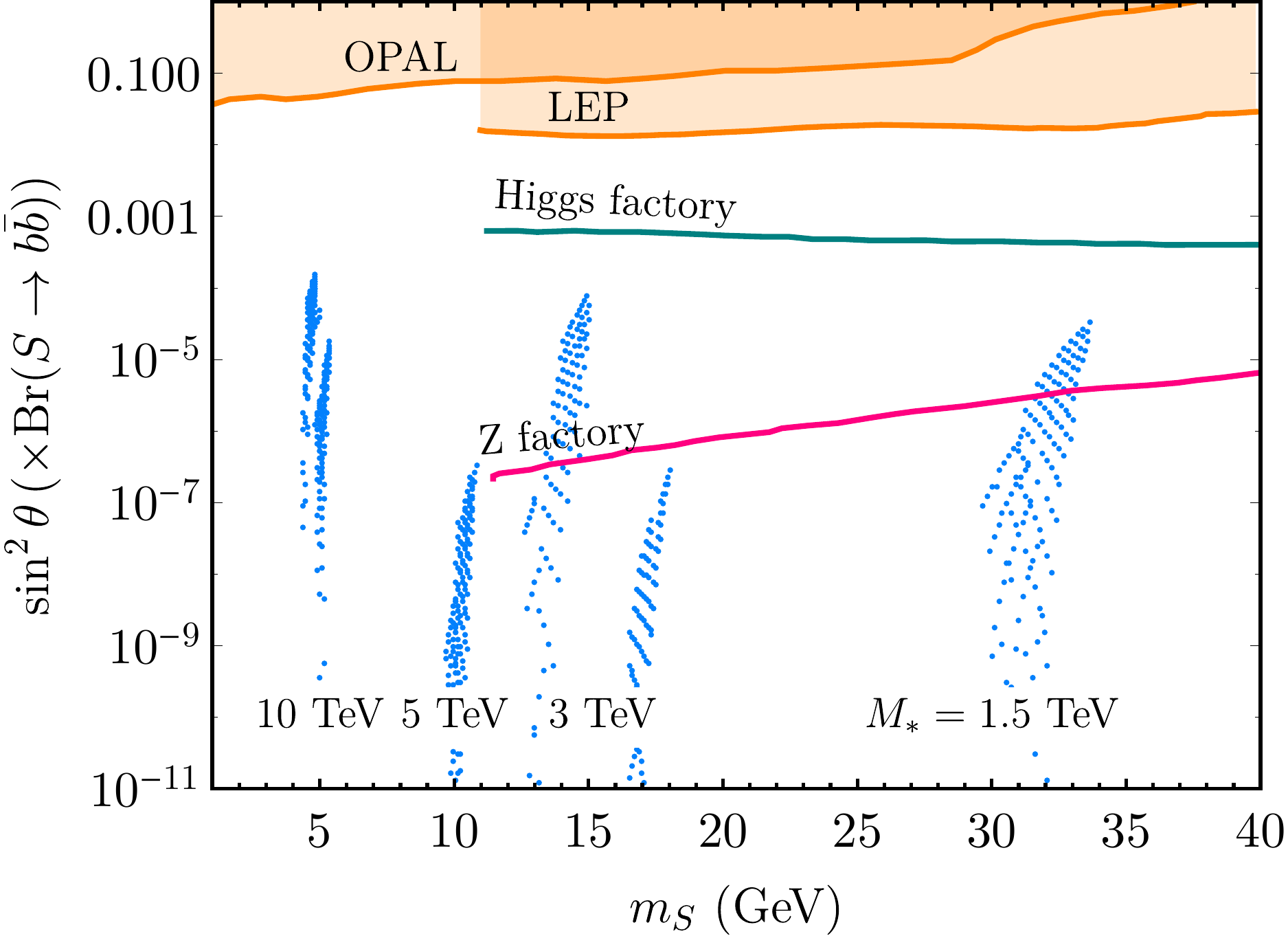}
    \caption{Benchmark points (blue dots) and various experimental constraints (lines with shading) 
    and prospects for future experiments (lines without shading) are shown
    in the plane of $m_s$-$\sin^2 \theta \times {\rm BR}(s \rightarrow b \bar{b})$. 
    LEP bounds are at the $95\%$ C.L. and Higgs/$Z$ prospects correspond to $3\sigma$ evidence
    curves~\cite{Chang:2017ynj}. Note that the benchmark points at $m_S\lesssim10~$GeV are not multiplied with the $b\bar{b}$ branching fraction.}
    \label{fig:exclusion}
\end{figure}

The dominant gluonic decay mode is an important difference between the singlet discussed here
and typical singlet scalar extensions of the Standard Model.  Nevertheless, although subdominant, the $b\bar{b}$ final state 
provides a useful search mechanism at particle colliders. The best limits for $m_s\simeq 10-100$~GeV are 
from Higgs searches at LEP~\cite{Barate:2003sz}, which probe $\sin^2\theta\gtrsim 0.01$. 
If the singlet is lighter than the $b\bar{b}$ threshold, the best constraint is from OPAL, 
which requires $\sin^2\theta \lesssim 0.1$ independently of its decay mode~\cite{Abbiendi:2002qp}. 
Smaller mixing angles, including the region of interest studied here for a light scalar, 
can be probed in the future at Higgs and/or $Z$ factories~\cite{Chang:2017ynj}. 
We show several current constraints and future searches in the plane of $m_s$ and the quantity
$\sin^2 \theta \times {\rm BR}(s \rightarrow b \bar{b})$
in Figure~\ref{fig:exclusion}, together with the points from our benchmark scans.

It is worth pointing out that the larger coupling to gluons may present interesting opportunities to search for low mass
scalar particles which are produced at high transverse momentum 
via the strong force, but decay through mixing with the Higgs into clean
final states such as into muons. 

\subsection{Gravitational Waves}

At the time of first confinement, $\Lambda_{\rm QCD} \sim \mathcal{O}(100\, \GeV)$, the EW
symmetry is unbroken. Thus, all of the SM quarks are light ($N_f=6$), and the resulting phase transition is expected to be 
first order. At the time of deconfinement, which happens shortly after confinement as the scalar fields roll to the SM-like vacuum, the 
top quark may have a mass comparable to the temperature, but for most relevant regions of parameter space all of the 
other quarks are light. Then, the first-order confinement phase transition is followed by a subsequent deconfining first-order phase transition, both occurring through bubble nucleation.  

As is well known, first-order phase transitions in the early Universe produce a stochastic background spectrum of gravitational 
waves (GWs), of a characteristic power-law form \cite{Hindmarsh:2017gnf,Cutting:2018tjt}. 
Contributions to this spectrum come from the collisions of the bubble walls themselves, and from the linear (acoustic) and 
non-linear (turbulent) dynamics in the plasma coupling to the bubble wall. 
Which contribution dominates is an open question for phase transitions in which no gauge 
bosons partake, and depends on the effective friction on the bubble wall by the plasma.

In the absence of a reliable effective field theory description of chiral symmetry breaking at the scales of interest, 
the gravitational wave spectra can be studied using the linear-sigma model as a low energy effective theory, 
with finite-temperature corrections from meson loops \cite{Bai:2018dxf,Croon:2019iuh,Helmboldt:2019pan} 
or using interpolating models such as the Nambu-Jona-Lasinio model \cite{Helmboldt:2019pan}. 
However, such models are known to fall short for the case of QCD, and the resulting GW spectrum is 
subject to very large uncertainties. In the present work we will therefore limit ourselves to some 
qualitative observations about the expected GW spectra, 
leaving a more detailed study of the GW phenomenology to future studies.

Sequential first-order phase transitions and the resulting GW spectra are a fascinating possibility 
which has not been explored in much detail. Such phenomenology has been suggested in the context of 
multi-step perturbative phase transitions \cite{Croon:2018new}, for example in an enhanced flavor 
sector \cite{Greljo:2019xan}. In the case studied in this paper, the scales of both phase transitions imply nucleation temperatures in the 
range $T_N = \mathcal{O}(10-100 \, \GeV)$. This implies that the resulting peak GW frequencies fall 
within the observational windows of space-based interferometer experiments such as LISA \cite{Caprini:2019egz}. 
However, as both phase transitions occur in the same sector, the plasma dynamics generated by the first phase transition is 
disrupted by the occurrance of the second transition, and one would typically expect a GW signal with a 
double peak, where the high frequency peak is lower in amplitude.

\section{Conclusions and Outlook}

We propose a simple model in which the SM is aided by an axion and a singlet scalar, which leads to a rich phenomenology based on a novel cosmological history,
realizing the observed baryon asymmetry by means of high-temperature QCD confinement 
and simultaneous electroweak symmetry breaking. 
We study the scalar potential in the confined phase, including mesonic (thermal) corrections to the Higgs potential. 
These corrections, along with couplings between the singlet and the SM Higgs, conspire to relax the model into a SM-like vacuum state before the onset of BBN.  It exemplifies how simple dynamics could result in radical changes to cosmology at high temperatures, and how such modifications may shed light on the mysteries of particle physics such as the baryon asymmetry of the Universe.

The hallmark of the dynamics is a light scalar particle whose mass is of order 10 GeV, with large coupling to gluons and relatively small mixing with the SM Higgs boson.  While not significantly constrained by current observations, future Higgs or $Z$ factories can probe some of the relevant parameter space which realizes the baryon asymmetry,
and optimized LHC searches for low mass particles, produced through strong interactions but decaying through Higgs mixing, could offer additional opportunities.

Moreover, the sequential phase transitions of QCD confinement and deconfinement potentially both occur through bubble nucleation, 
and therefore, may give rise to a characteristic doubly-peaked gravitational wave spectrum. 
As these transitions take place at temperatures $T_N \sim \mathcal{O}(10-100 \GeV)$, the resulting stochastic 
background would be strongest in the frequency bands of space-based interferometers. 
A simplified version of this model could in principle be studied in finite temperature lattice gauge theory. 
Such a study could determine the order of the confining phase transition, and be used to 
estimate the resulting gravitational wave spectrum. 

This paper leaves several interesting questions for future research. For example, the axion relic abundance may be affected by the evolution of the confinement scales. One may also be interested in UV-completions of the current model. The analysis in this paper applies to an effective theory at low energies, which may be generated in by fluctuations of a radion or dilaton field in an extra-dimensional model or through vector-like quarks at $\sim$TeV scales.

A mechanism in which confinement triggers subsequent dynamics, such as studied in this paper, could also be employed in other contexts. For example, confinement may occur at lower temperatures, such that a period of supercooling ensues while the scalar field is stuck in the confining vacuum. 
Then, for $V(v_s)>\rho_{\rm rad}(\Lambda_{\rm QCD})$ this vacuum will start to inflate until the confinement scale is reached. Hence, the succession of steps that confinement sets in motion implies a graceful exit to a brief period of late inflation.

\section*{Acknowledgements}
The authors thank Juan Garcia-Bellido, Belen Gavela, David McKeen, Veronica Sanz, Jose Miguel No and Tevong You for useful discussions. 
TMPT and SI thank Haolin Li for collaboration on related topiocs.
This work was supported in part by the NSF via grant number PHY-1915005 and DGE-1839285. TRIUMF receives federal funding via  a  contribution  agreement  with  the  National  Research  Council  of  Canada  and  the  Natural Science and Engineering Research Council of Canada. SI acknowledges support from the University Office of the President via a UC Presidential Postdoctoral fellowship. This work was partly performed at the Aspen Center for Physics, which is supported by NSF grant PHY-1607611.

\appendix

\section{Scalar Mixing}
\label{sec:smixing}

The mass eigenstates and mixing angles are obtained by 
diagonalizing the mass-squared matrix at zero temperature:
\begin{align}
 M^2 &= \begin{pmatrix} 
m^2_{h,h} & m^2_{h,s} \\
m^2_{h,s} & m^2_{s,s} 
\end{pmatrix} = \begin{pmatrix} 
\partial^2 V/ \partial v_h^2 & \partial^2 V/ \partial v_s \partial v_h \\
\partial^2 V/ \partial v_s \partial v_h & \partial^2 V/ \partial v_s^2 
\end{pmatrix} \\
&= \begin{pmatrix}
    2\lambda_h v_h^2 && -\frac{b_1}{\sqrt{2}}v_h +b_2 v_s v_h \\
    -\frac{b_1}{\sqrt{2}}v_h +b_2 v_s v_h && -\frac{a_1}{\sqrt{2}v_s}+\frac{3a_3}{2\sqrt{2}} v_s + 2 a_4 v_s^2 + \frac{b_1}{2\sqrt{2}}\frac{v_h^2}{v_s}
    \end{pmatrix} \label{eq:masssquare}
\end{align}
where $v_s$ and $v_h$ should be understood to be their zero temperature values, which are assumed to be non-zero.  We have also invoked the conditions obtained by minimizing $V$ to obtain the nice form of \ref{eq:masssquare}.

Determining the mass eigenstates is accomplished by
finding the orthogonal matrix $O$, such that $M^2 = O^T M^2_{\rm diag} O$. Without loss of generality we choose 
\begin{align}
O = \begin{pmatrix} 
\cos \theta & - \sin \theta \\
\sin \theta & \cos \theta 
\end{pmatrix} 
\end{align}
where $\theta$ is the mixing angle between the new scalar and the SM Higgs. Therefore, 
\begin{align}
M^2 &= \begin{pmatrix} 
\cos \theta & \sin \theta \\
- \sin \theta & \cos \theta 
\end{pmatrix} \begin{pmatrix} 
M^2_{h} & 0 \\
0 & M^2_{S} 
\end{pmatrix} \begin{pmatrix} 
\cos \theta & - \sin \theta \\
\sin \theta & \cos \theta 
\end{pmatrix} = O^T M^2_{\rm diag} O 
\end{align}
where $M^2_h$ and $M^2_S$ are the eigenvalues of $M^2$:
\begin{align}
M_h^2 &= \frac{(m^2_{h,h}+m^2_{s,s}) + \sqrt{(m^2_{h,h}-m^2_{s,s})^2 + 4(m^2_{h,s})^2}}{2} \\
M_s^2 &= \frac{(m^2_{h,h}+m^2_{s,s}) - \sqrt{(m^2_{h,h}-m^2_{s,s})^2 + 4(m^2_{h,s})^2}}{2}
\end{align}
$M^2_h$ is chosen to be the SM Higgs' mass which means that we have implicitly assumed that $M_S$<$M_h$ by choosing $M_h^2$ to be the larger eigenvalue. This equation then implies that $\theta$ must satisfy
\begin{align}
m^2_{h,h} &= M_h^2 \cos^2 \theta + M_s^2 \sin^2 \theta\\
m^2_{s,s} &= M_h^2 \sin^2 \theta + M_s^2 \cos^2 \theta\\
m^2_{h,s} &= (-M_h^2 + M_s^2)\sin \theta \cos \theta
\end{align}
We subtract the first from the second of these equations and use a trigonometric identity to obtain an expression for $\cos 2\theta$. Similarly, we use the third equation and a trigonometric identity to find $\sin 2\theta$. 
\begin{align}
\cos 2\theta &= \frac{m^2_{h,h} - m^2_{s,s}}{(M_h^2 - M_s^2 )} \\
\sin 2\theta &= \frac{-2 m^2_{h,s} }{(M_h^2 - M_s^2)} 
\end{align}
The code which calculates the mixing angle as a function of $M_S$ for the benchmarks discussed in this paper can be found here: \href{https://github.com/jnhoward/EarlyQCD}{https://github.com/jnhoward/EarlyQCD}.

\bibliographystyle{JHEP}
\bibliography{references}
\end{document}